**Excitation of multiple 2-mode parametric resonances by a single driven mode**

Authors: Adarsh Ganesan[1], Cuong Do[1], Ashwin Seshia[1]

1. Nanoscience Centre, Department of Engineering, University of Cambridge, Cambridge, UK

**We demonstrate autoparametric excitation of two distinct sub-harmonic mechanical modes by the same driven mechanical mode corresponding to different drive frequencies within its resonance dispersion band. This experimental observation is used to motivate a more general physical picture wherein multiple mechanical modes could be excited by the same driven primary mode within the same device as long as the frequency spacing between the sub-harmonic modes is less than half the dispersion bandwidth of the driven primary mode. The excitation of both modes is seen to be threshold-dependent and a parametric back-action is observed impacting on the response of the driven primary mode. Motivated by this experimental observation, modified dynamical equations specifying 2-mode auto-parametric excitation for such systems are presented.**

Parametric resonance has been widely studied in many physical systems, including in micromechanical resonators [1-10]. Parametric resonance provides an interesting route for indirect excitation of tones of frequencies lower than that of drive tone. The signal amplification associated with this phenomenon allows for the improvement in oscillator metrics and has been applied to the design of highly sensitive sensors [11-13]. Autoparametric resonance is associated with self-excitation of a sub-harmonic mode mediated by non-dissipative coupling to the driven mode. These mechanical modes can be such that they are excited through the internal coupling to the drive mode but might not be accessible to direct driving using conventional transducer arrangements. We have previously demonstrated 2-mode [10] and 3-mode [9] auto-parametric excitation in microelectromechanical structures. This letter demonstrates the parametric excitation of different sub-harmonic mechanical modes using the same driven mode actuated at different drive frequencies.

Autoparametric resonance is usually mathematically modelled by coupled oscillators. Here, the coupling is established between the harmonic oscillators of frequency ratio 1:2. The assumed deformation modes of micromechanical structure are represented as $u_{m1}(\vec{r},t) = f(\vec{r})u_{m10}(t)$ and $u_{m2}(x,t) = g(\vec{r})u_{m20}(t)$. At a particular observation location, the dynamics of nonlinearly interacting damped harmonic oscillators ($Q_{i=1,2}$) can then be modelled as

$$\ddot{Q}_1 = -\omega_1^2 Q_1 + P\cos\Omega t + \mu Q_2^2 + \chi_{12} Q_1 Q_2 - 2\zeta_1 \omega_1 \dot{Q}_1 \qquad (1\text{-}1)$$

$$\ddot{Q}_2 = -\omega_2^2 Q_2 + \chi_{21} Q_1 Q_2 - 2\zeta_2 \omega_2 \dot{Q}_2 \quad (1\text{-}2)$$

Here, $\omega_{i=1,2}$ and $\zeta_{i=1,2}$ are the natural frequencies and damping coefficients of harmonic oscillators $Q_{i=1,2}$, $\chi$ is the coupling coefficient between the modes $Q_1$ and $Q_2$, $\mu$ is the self-coupling coefficient of mode $Q_2$, $P$ and $\Omega$ are the amplitude and frequency of drive. When the drive frequency $\Omega$ is close to the resonant frequency of oscillator $\omega_1$, at high enough drive levels $P$, a parametric load in the harmonic oscillator $Q_2$ is introduced through the coefficient $\chi$ as provided in the equation (1-2). This loading causes parametric instabilities of type $\Omega \cong \omega_1 \cong 2\omega_2$. This framework however, considers the case where only a single sub-harmonic oscillator $Q_2$ can be auto-parametrically excited. It is also seen that a parametric back-action exists where the response of mode, $Q_1$, can in turn be modulated by the excitation of mode $Q_2$.

In this letter, we hypothesize a more general case where a harmonic oscillator $Q_1$ can independently trigger different harmonic oscillators $Q_{n+1}; n = 1,2,3,...$ at different drive frequencies within the resonant dispersion band of harmonic oscillator $Q_1$. The particular case of 2 sub-harmonic modes is shown in the eq. (2).

$$\begin{aligned}
\ddot{Q}_1 &= -\omega_1^2 Q_1 + P \cos \Omega t + \sum_{i=2}^{n+1} \mu_i Q_i^2 + \sum_{i=2}^{n+1} \chi_{1i} Q_1 Q_i - 2\zeta_1 \omega_1 \dot{Q}_1 \\
\ddot{Q}_2 &= -\omega_2^2 Q_2 + \chi_{21} Q_1 Q_2 - 2\zeta_2 \omega_2 \dot{Q}_2 \\
\ddot{Q}_3 &= -\omega_3^2 Q_2 + \chi_{31} Q_1 Q_3 - 2\zeta_3 \omega_3 \dot{Q}_3
\end{aligned} \quad (2)$$

This physical picture is experimentally observed in a piezoelectric-on-silicon micromechanical device based on a free-free beam topology of dimensions 1100 $\mu m$ X 350 $\mu m$ X 11 $\mu m$ (Figure 1(a)). The electrical signals are fed to the device through 1 $\mu m$ thick Al electrodes and the piezoelectric excitation of micromachined structure is established through 0.5 $\mu m$ thick AlN layer. The mechanical motion of the device is monitored by Laser Doppler Vibrometry with a spatial lateral resolution of 35 $\mu m$ corresponding to the laser spot size. A length extensional mode of frequency $\sim 3.86\ MHz$ (Figure 1(b)) has been identified in this device and is considered for demonstrating auto-parametric excitation. At high drive levels of this mode, the self-excitation of a subharmonic mode results due to autoparametric instability (Figure 1(c)). There is an intrinsic parametric gain associated with this excitation. When the drive frequency is $3.862\ MHz$, mode 1 gets excited (Figure 1(d)). However, when the drive frequency is $3.876\ MHz$, mode 2 gets excited (Figure 1(e)). Hence, as experimentally observed, it is possible for multiple confined modes to be parametrically excited using the same driven mode.

To further quantitatively describe this behaviour, experiments were carried out at different drive conditions. At first, keeping the drive frequency constant at $\Omega = 3.852\ MHz$, the power level $S_{in}$ is varied from $-4$ to $22\ dBm$. Figure 2(a) shows that there is an inherent threshold associated with this self-excitation process. Upon crossing this threshold $T = 6\ dBm$, the self-excitation of sub-harmonic frequency component takes place. Corresponding to this self-excitation, an increase in the displacement amplitude of driven mode is observed (Figure 2(b)) This is modelled through the quadratic nonlinear terms $Q_i^2$ & $Q_1 Q_i$; $i = 2,3,...$ in driven harmonic oscillator. Then, to obtain the drive frequency dependence on the parametric excitation threshold $T$, the drive frequency $\Omega$ has been tuned from 3.836 to 3.888 $MHz$. The parametric excitation threshold $T$ associated with each drive frequency (Figure 2(c)) appears to be associated with the Lorentzian decay of the driven mode amplitude ($Q_1$) (Figure 2(d)). However, the displacement profiles corresponding to the sub-harmonic frequencies are not same for all drive frequencies (Figures 3(a-c) and supplementary section). This can be explained by the physical picture of simultaneous parametric interactions between $Q_1$ and $Q_i$; $i = 2,3,...,n + 1$. That being said, at sub-harmonic frequencies, a combination of multiple sub-harmonic oscillators $Q_i$; $i = 2,3,...,n + 1$ gets excited. However, corresponding to the resonant frequency $\omega_i$ and damping coefficient $\zeta_i$ of each oscillator, the resultant mode-shapes are evaluated at different frequencies. In the frequency range $3.836 - 3.872\ MHz$, due to maximal strength associated with mode 1, the resultant mode shape corresponds to mode 1 (Figures 3(a) and supplementary section). Similarly in the frequency range $3.876 - 3.888\ MHz$, the resultant mode shape corresponds to mode 2 (Figures 3(b) and supplementary section). However, at the cross-over frequency $3.874\ MHz$, a combination of mode shapes corresponding to modes 1 and 2 is observed (Figures 3(c) and supplementary section).

In summary, this paper highlights the possibility of a generic physical picture of simultaneous interactions between multiple mechanical modes in auto-parametric resonance. We have proved this physical picture by demonstrating the excitation of two sub-harmonic modes $Q_2, Q_3$ using $Q_1$. However, this mechanism can be extended to more than two confined sub-harmonic modes. The relevance of this physical picture could be extended to other physical systems undergoing 2-mode parametric instability. The demonstration of this previously unexplored pathway in auto-parametric resonance could be applied to the selective excitation and detection of internal modes with potential application to inertial imaging [14] and frequency-selective mechanical signal processing [15].

**Acknowledgments**

Funding from the Cambridge Trusts is gratefully acknowledged.

**Authors' contributions**

AG and CD designed the device and performed the experiments; AG analysed the results and wrote the manuscript; AAS supervised the research.

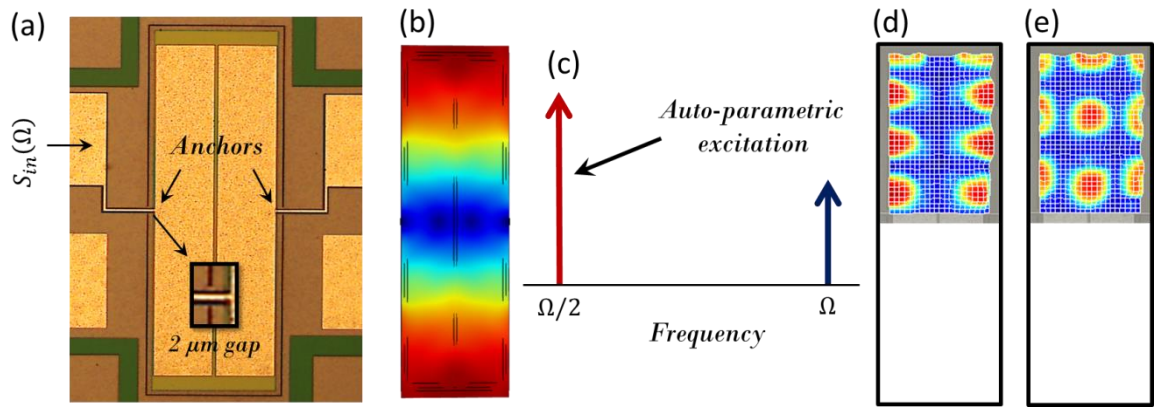

Figure 1: **Observation of 2-mode autoparametric excitation: (a)**: Signal $S_{in}(\Omega)$ is applied on a free-free beam microstructure; **(b)**: The length extensional mode shape (eigenfrequency $\sim 3.86\ MHz$); **(c)**: Auto-parametric excitation of the tone at $\Omega/2$; The 2-D displacement profiles at $\Omega/2$ when **(d)**: $\Omega = 3.862\ MHz$; **(e)**: $\Omega = 3.876\ MHz$. Note: The displacement profiles are normalized to the maximum displacement in the structure- i.e. Red corresponds to 1 and blue corresponds to 0.

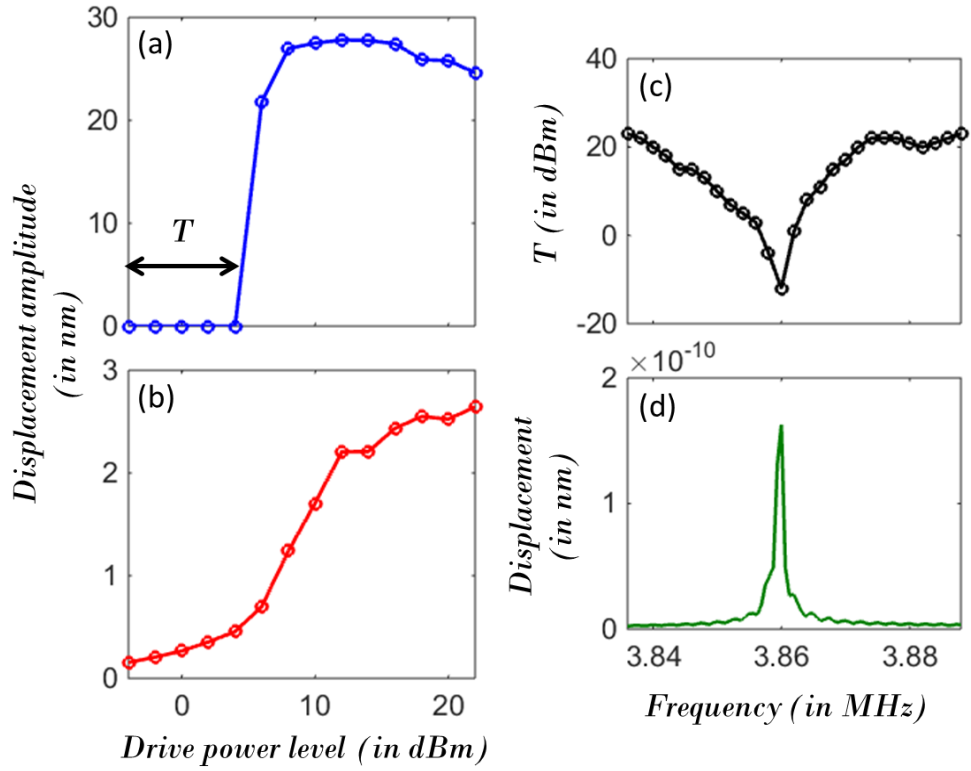

Figure 2: **Parametric excitation threshold:** The displacement amplitudes at **(a)**: $\Omega/2$; **(b)**: $\Omega$ for $S_{in}(\Omega = 3.852\ MHz) = -4 - 22\ dBm$; **(c)**: The parametric excitation threshold $T$ for drive frequencies $3.836 - 3.872\ MHz$; **(d)**: The linear resonant dispersion curve which is obtained at the drive power level of $-20\ dBm$.

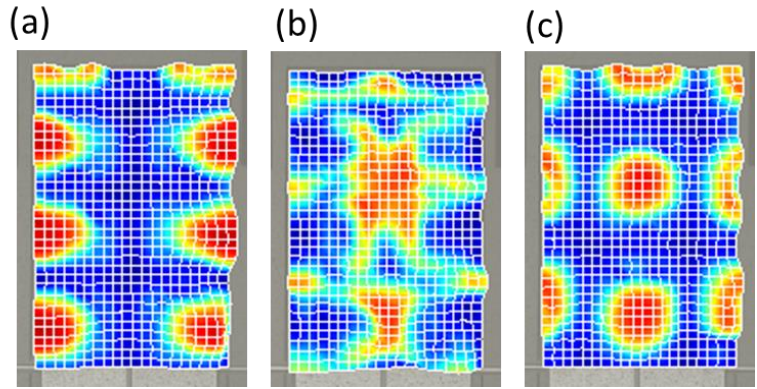

Figure 3: **Excitation of different sub-harmonic modes:** The 2-D displacement profiles at $\Omega/2$ when **(a)**: $\Omega = 3.862\ MHz$; **(b)**: $\Omega = 3.874\ MHz$; **(e)**: $\Omega = 3.876\ MHz$. Note: The displacement profiles are normalized to the maximum displacement in the structure - i.e. Red corresponds to 1 and blue corresponds to 0.

.